\documentclass{appolb}
\usepackage{graphicx} 
\usepackage{appendix}
\usepackage{xcolor}
\usepackage[backend=biber,style=numeric,sorting=none]{biblatex}
\def\Cpg {$^{12}$C(p,$\gamma$)$^{13}$N }
\def\Cpn {$^{13}$C(p,n)$^{13}$N }
\def\b+ {$\beta^+$ }
\def\C12 {$^{12}$C }
\def\C13 {$^{13}$C }
\def\N13 {$^{13}$N }

\title{Influence of \C13 on proton-induced \N13 production from natural carbon target}
\author{Tomasz Matulewicz 
and Izabela Skwira-Chalot
\address{Institute of Experimental Physics, Faculty of Physics, University of Warsaw}}
\date{October 2025}

\begin{document}

\maketitle

\begin{abstract}
Two recent measurements of \Cpg reaction on natural carbon were performed by detecting \b+ decay of \N13 residue.
It is argued that the measurements at energies above the \Cpn reaction threshold of 3.24 MeV proton kinetic energy can be interpreted as a consequence of 1.06\% admixture of \C13 isotope in natural carbon, as the cross section for \N13 production raised by 3 orders of magnitude with respect to the measurements of \Cpg at lower energies.
\end{abstract}

\section{Introduction}

The radiative capture reaction \Cpg has been recently studied via the detection of \b+ decay of \N13 isotope. 
These measurements were performed on natural carbon targets in the astrophysical proton energy range below 2 MeV \cite{ATOMKI2023radiative} and in the range 2-22 MeV \cite{rodriguez2022production}. 
The cross sections measured by Rodr{\'\i}guez-Gonz{\'a}lez et al. \cite{rodriguez2022production} at proton energies 2-22 MeV are three orders of magnitude above that reported in measurements performed at ATOMKI \cite{ATOMKI2023radiative}. 
We argue that this unexpected increase might be due to the (p,n) reaction on \C13 nuclei, since the target was natural carbon.
The cross section of \Cpn reaction leading to the formation of \N13 is relatively high, what might compensate the small 1.06\%
\cite{IUPAC2015} content of \C13 in natural carbon.

\section{Analysis}

The \N13 nucleus can be produced in two proton-induced reaction channels on natural carbon, namely \Cpg and \Cpn . 
The first one on the most abundant $^{12}$C
isotope (about 99\% \cite{IUPAC2015}) is exothermic.
The second one on the $f_{13}$=1.06\% abundant $^{13}C$ \cite{IUPAC2015}, is endothermic with \\ Q=$\left(M(^{13}C)+M(p)-M(^{13}N)-M(n) \right)c^2=-$3.003 MeV. 
The corresponding proton beam energy threshold is E$_p^{thr}$=3.236 MeV.
The cross section of the radiative proton capture reaction is reduced by the electromagnetic vertex.
This is not the case of the $(p,n)$ reaction, where all vertices are strongly coupled.
The natural carbon target for protons below E$_p^{thr}$ can be treated as monoisotopic, while at higher energies the (p,n) reaction may start to play a role, particularly as the cross section is significantly higher. 
This factor might compensate the low abundance of \C13 isotope.

Recently, two experimental groups measured the \N13 production via the detection of \b+ decay \cite{ATOMKI2023radiative, rodriguez2022production}.
Both groups used natural carbon targets and measured the decay curve of \N13 by the detection of 511 keV $\gamma$ quanta from the \b+ annihilation.
Single HPGe detector was used to detect the off-beam 511 keV radiation in the ATOMKI measurements \cite{ATOMKI2023radiative}, while a PET scanner was used off-beam to detect coincident events in the measurements by Rodr{\'\i}guez-Gonz{\'a}lez et al. \cite{rodriguez2022production}.
The half life of \N13 is equal to $T_{1/2}=9.951\pm 0.003$ min. \cite{long2022_13Nlifetime}, making it very distinct from other $\beta^+$-active nuclei of similar mass - half life of $^{11}$C is twice longer, while half life of $^{15}$O is five times shorter (unexpected presence of oxygen in a BN sample irradiated with protons was observed this way \cite{matulewicz2024nitrogen}).
The measurement of \Cpg reaction at astrophysically important energy range below 2 MeV was performed at ATOMKI \cite{ATOMKI2023radiative}.
The results obtained by detecting the \N13 decay agree well with the measurements employing other methods.
As the energy range in ATOMKI measurements \cite{ATOMKI2023radiative} is below the threshold energy for \Cpn reaction, the process is only the radiative capture.
However, in the case of measurements by Rodr{\'\i}guez-Gonz{\'a}lez et al. \cite{rodriguez2022production} the proton energies were up to 22 MeV, which is well above E$_p^{thr}$.
Comparison of the cross section values reported for three highest proton energies
of \cite{ATOMKI2023radiative} and three lowest energies of Rodr{\'\i}guez-Gonz{\'a}lez et al. \cite{rodriguez2022production},
summarized in the Table \ref{tab:reactions}, shows a sudden rise of three orders of magnitude, from $\mu$b to mb range.

\begin{table}[htb]
\centering
\caption{Cross section of \Cpg reaction on natural carbon targets obtained by the detection of \b+ decay of \N13 . Three results at highest energies of  ATOMKI measurements \cite{ATOMKI2023radiative} and three results at lowest energies from \cite{rodriguez2022production} are listed.} 
\begin{tabular}{ccc}
 Energy (MeV) & cross section  & ref. \\ \hline
 1.7977 & (2.98 $\pm$ 0.25) $\mu$b &  \\ 
1.8478 & (1.85 $\pm$ 0.15) $\mu$b & \cite{ATOMKI2023radiative} \\ 
1.898 & (1.54 $\pm$ 0.13) $\mu$b &   \\  \hline
2.0 $\pm$ 1.2 & (0.79 $\pm$ 0.06) mb  &    \\ 
4.2 $\pm$ 1.0  & (1.89 $\pm$ 0.15) mb & \cite{rodriguez2022production}   \\ 
5.9 $\pm$ 0.8  & (2.09 $\pm$ 0.17) mb  &    \\ 
\end{tabular}
    \label{tab:reactions}
\end{table}

We argue that the results of Rodr{\'\i}guez-Gonz{\'a}lez at al. \cite{rodriguez2022production} are indeed the cross section for the \N13 production on natural carbon, but the responsible reaction is mainly \Cpn and not \Cpg .
This conjecture is based on the experimental results of the \Cpn reaction, performed on isotopically enriched targets \cite{kitwanga1989production, firouzbakht1991measurement}.
The results of Firouzbakhtl et al. \cite{firouzbakht1991measurement} were obtained using a well-type NaI(Tl) scintillating detector, a method different from Kitwanga et al., employing coincident measurement in a pair of plastic scintillation detectors \cite{kitwanga1989production}.
These results are compared to the measurements of Rodr{\'\i}guez-Gonz{\'a}lez et al. \cite{rodriguez2022production} scaled by the $\frac{1}{f_{13}}$, as the correction of the abundance of $^{13}C$ in natural carbon. 
The results, shown in Fig. 1, agree quite well, particularly around the broad maximum in the proton energy range of 5-10 MeV.

\begin{figure}[htp]
    \centering
    \includegraphics[width=\textwidth]{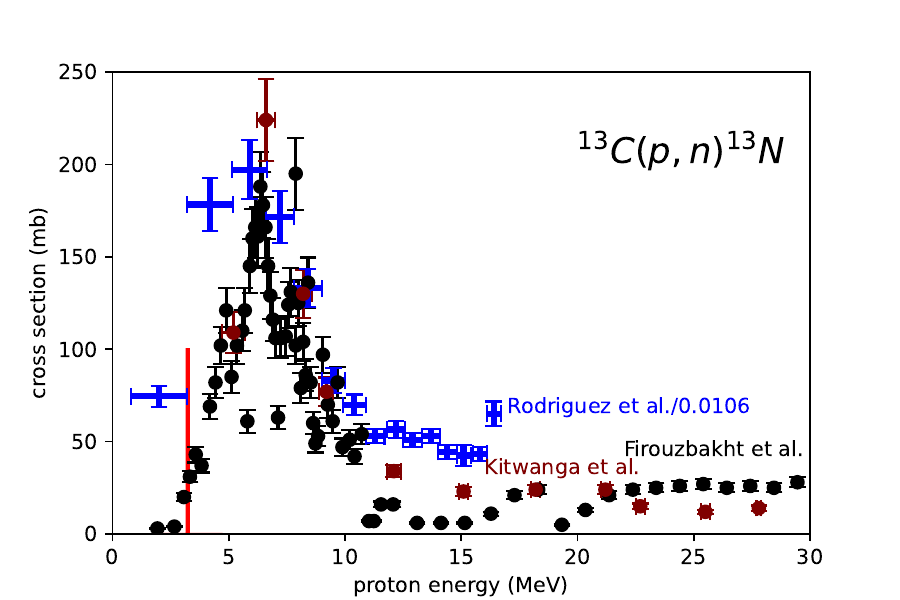}
    \caption{The cross section of \Cpn reaction from the measurements with isotopically enriched \C13 targets \cite{kitwanga1989production, firouzbakht1991measurement}. 
    The production of \N13 observed in the measurements of Rodr{\'\i}guez-Gonz{\'a}lez et al. \cite{rodriguez2022production} is scaled with the 1.06\% abundance of \C13 nuclei in the natural carbon target.}
    \label{fig:detection}
\end{figure}

The interpretation by Rodr{\'\i}guez-Gonz{\'a}lez et al. \cite{rodriguez2022production} which attributes the measurements to the radiative process \Cpg is, however, supported by the measurements of Cohen \cite{Cohen1955} performed on isotopically enriched $^{12}$C target at two proton beam energies of 5 and 11 MeV (the reported cross section is 2.5 mb and 1.8 mb, respectively).
The carbon target used in measurements of Cohen, performed more than 70 years ago, was isotopically enriched (\C13 content was as low as 0.05\%).
The influence of the (p,n) reaction on the remaining \C13 nuclei was accounted to be $\sim$10\%, however the total uncertainty of the measured cross section was not provided \cite{Cohen1955}.
It should be noticed, that other results of the measurements of (p,$\gamma$) by Cohen on $^{54}$Fe, $^{60}$Ni, $^{64}$Zn, and $^{209}$Bi \cite{Cohen1955} are in relatively good agreement with later experiments.

\section{Conclusions}

The measurement of the \Cpg reaction in the proton energy range between 2 MeV and 22 MeV, performed by Rodr{\'\i}guez-Gonz{\'a}lez et al. \cite{rodriguez2022production}, can be interpreted as the production of \N13 on the 1.06\% abundant \C13 isotope. 
The reported cross section, when corrected for low abundance of \C13 isotope, agrees roughly with the dedicated measurements of \Cpn reaction on isotopically enriched targets.

The importance of the cross section of proton-induced reactions on carbon, nitrogen and oxygen targets calls for an intensified experimental activity, as these nuclei are abundant in tissue and proton-therapy procedures cause nuclear reactions on these elements
(e.g. reactions on oxygen \cite{matulewicz2024oxygen}). 
This activity should cover not only the most abundant light isotopes of these elements, but also all heavier ones, as it is the topic of current astrophysical research \cite{LUNA2025CNO}.
The recent progress in experimental capabilities, in particular radiation detectors of energy resolution superior to those used in the past, should be exploited.
The results might influence as well the precision of predicted radiation in proton therapy centers, an issue valid for patients and personnel \cite{dose}.

\printbibliography 

\end{document}